\documentclass[a4paper]{jpconf}
\usepackage{graphicx}
\newcommand{\vk}{{\bf k}}

\newcommand{\vr}{{\bf r}}
\begin{document}
\title{Theory of Andreev reflection in a two-orbital model of iron-pnictide superconductors}

\author{M. A. N. Ara\'ujo$^{1,2}$ and P.  D. Sacramento$^{1}$}

\address{$^1$ CFIF, Instituto Superior 
T\'ecnico, UTL, Av. Rovisco Pais, 1049-001 Lisboa, Portugal}
\address{$^2$ Departamento de F\'{\i}sica,  Universidade de \'Evora, P-7000-671, \'Evora, Portugal}
\ead{mana@cfif.ist.utl.pt}

\begin{abstract}
A recently developed theory for the problem of Andreev reflection between a normal metal (N) 
and a multiband superconductor (MBS) assumes that the incident wave from the normal metal is coherently transmitted through several bands inside the superconductor. Such splitting of the probability amplitude into several channels is the analogue of a quantum waveguide. 
Thus, the appropriate matching conditions for the wave function at the N/MBS interface are derived from an  extension of quantum waveguide theory. 
Interference effects between  the transmitted waves inside the superconductor  manifest themselves in the conductance.  
We provide results for a FeAs superconductor, in the framework of a recently proposed  effective two-band model and two recently proposed gap symmetries: in the  sign-reversed  s-wave 
($\Delta\cos(k_x)\cos(k_y)$) scenario resonant transmission through surface  Andreev bound states (ABS) at nonzero energy is found as well as destructive interference effects  that produce zeros in the conductance; in the  extended s-wave ($\Delta\left[\cos(k_x)+\cos(k_y)\right]$) scenario no ABS at finite energy are found.   
\end{abstract}

\section{Introduction}
Electronic scattering at the interface between a normal metal 
and a superconductor has been  used as a probe to investigate the electronic properties of FeAs superconductors (FAS)  \cite{tesanovic,chineses}.
FAS have a complex band structure, 
with a Fermi surface (FS) consisting of four sheets, 
two of them hole-like 
(labeled as $\alpha_1$,$\alpha_2$), and the other two electron-like (labeled as $\beta_1$,$\beta_2$)  \cite{singh,xu,mazin,haule}.
S-wave, d-wave and p-wave pairing scenarios have been proposed to describe the
superconducting state  \cite{chubukovs,qi,lee,yao}, as well as extended s-wave
 \cite{graser}.

Recently, two of the suggested pairing scenarios have been in dispute  \cite{mazin,graser}.  
In the "sign-reversed s-wave state" (s$^\pm$-state), the gap function 
has no nodes on any of the FS sheets but takes on
opposite signs on the hole-like and the electron-like sheets of the FS.
Throughout the unfolded BZ, the gap function can be approximated as
$\Delta(\vk)=\Delta\cos(k_x)\cos(k_y)$.  
The other pairing symmetry is the extended s-wave, in which 
the gap function has four nodes on the electron FS sheets, no nodes on the hole sheets,
and has the same sign on both hole FS's.
We shall approximate the gap function as 
$\Delta(\vk)=\Delta\left[\cos(k_x)+\cos(k_y)\right]$. 
On the experimental side, the results for lanthanum based superconductors (LaFeAsO family) seem to be mostly consistent with unconventional gap functions with nodes on the FS, while the experiments with superconductors containing  rare-earth atoms ([Re]FeAsO family) as well as the Fe$_2$As$_2$
family tend to suggest a nodeless nearly isotropic gap across the FS.  
We note, however, that Andreev reflection studies have lead
to different conclusions on the  pairing symmetry for 
the same (rare-earth based) material \cite{tesanovic,chineses}.

Existing theories of Andreev scattering \cite{blonder,tanaka} apply to one band  superconductors
and, in the analysis of experimental data for multiband superconductors, the 
bands are usually treated as separate conduction channels with (classically) additive conductances. 
Such theories neglect the quantum mechanical nature of  the scattering problem at the interface, where interference effects between the transmission channels (bands) inside the MBS are expected.

\section{Matching conditions at the N/MBS interface}

\begin{figure}[h]
\includegraphics[width=16pc]{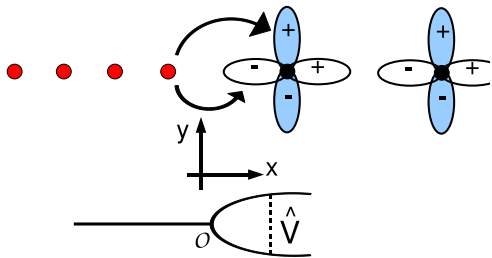}\hspace{3pc}%
\begin{minipage}[b]{14pc}
\caption{\label{tightexample}The analogy between the N/MBS interface (top) and a quantum waveguide with three branches joining at the circuit node $O$ (bottom). The hibridization, $\hat V$, between branches in the latter simulate the hopping between d$_{xz}$ and d$_{yz}$ orbitals in the former.}
\end{minipage}
\end{figure}

We recently addressed  \cite{us} the problem of interface scattering between a one-band metal  a MBS.
A two-orbital model that reproduces the FS's of iron pnictides assumes two orbitals on each site, the d$_{xz}$ and the d$_{yz}$ orbitals
(although a larger number of bands has also been argued necessary  \cite{cvetkovic}).
A one-dimensional model of the  N/MBS interface is a tight-binding  chain 
with a bifurcation because the electron on the normal metal atom that is 
closest to the interface can hopp to either one of the two Fe orbitals (see Figure \ref{tightexample}). 
This is the same quantum mechanical problem as in a quantum waveguide. 
In the two dimensional case, the tight-binding chain is repeated along the $y$ (interface) direction. 
The appropriate matching conditions for the wave function at the interface then follow  as an  extension of quantum waveguide theory to this  problem  \cite{us}. 
While an incoming  plane wave state in the normal metal, with energy $\varepsilon(\vk)$, takes the simple form $e^{i\vk\cdot\vr}$, 
a plane wave in the MBS has a two-component form:
\begin{equation}
\phi_\vk(x>0) = \left(\begin{array}{c}\alpha_\vk \\ \beta_\vk\end{array}\right)
e^{i\vk\cdot\vr}\,,\label{planwav}
\end{equation}
where $\alpha_\vk$ and $\beta_\vk$ denote projections on the d$_{xz}$ and d$_{yz}$ branches, respectively, 
and are obtained from the two orbital model Hamiltonian, $\hat H_S$, 
for the FAS that has been described in Ref \cite{raghu}. 
In the superconducting state, we assume that an elementary excitation is a linear combination of plane waves with opposite momenta with Bogolubov-deGennes amplitudes $u$ and $v$.  
The transmitted part of the wave function, $\psi(x>0)$, is a superposition of plane waves of the form (\ref{planwav}).
The first matching condition establishes that the wave function must be single-valued at the node $O$ where the three branches meet:
\begin{eqnarray}
\psi(x=0^-)=(1,0)\cdot \psi(x=0^+) = (0,1)\cdot \psi(x=0^+)\,.\label{kirkof1}
\end{eqnarray} 
The probability current conservation at the circuit node is garanteed by the second matching condition:
\begin{eqnarray}\left[
\frac{\partial\varepsilon}{\partial \hat k_x}\ \psi\right]_{x=0^-} = 
(1,1)\cdot\left[\frac{\partial\hat H_S}{\partial \hat k_x}\ \psi\right]_{x= 0^+}\,,
\label{kirkof2}
\end{eqnarray}
where the Hamiltonian is written in momentum space and  the operator $\hat \vk=-i\nabla$ in the continuum limit. The left multiplication by $(1,1)$ gives the sum of the transmitted currents through branches d$_{xz}$ and d$_{yz}$, which is equal to the incoming current from the normal metal.

\section{Results}

\subsection{s$^\pm$-wave gap symmetry}

\begin{figure}[t]
\includegraphics[width=20pc]{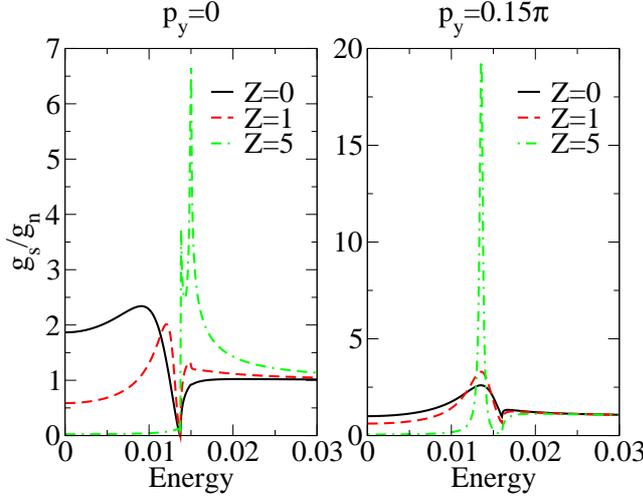}
\begin{minipage}[b]{18pc}
\caption{\label{gsgn_swave_Z}Relative conductance, $g_s/g_n$, for $s^{\pm}$-wave symmetry,
for different 
disorder values, $Z$, at normal incidence ($p_y=0)$ and for a finite transverse momentum, $p_y$.
Destructive interference effects between the transmitted plane waves produce 
a zero in the conductance at normal incidence.}
\end{minipage}
\end{figure}

\begin{figure}[h]
\vspace{1pc}
\includegraphics[width=18pc]{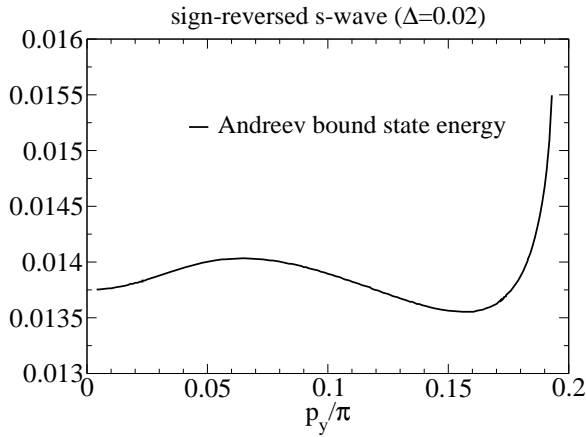}\hspace{2pc}%
\begin{minipage}[b]{18pc}
\caption{\label{ABSfigure}Energy of the surface Andreev bound state as function of the 
transverse momentum, for $s^{\pm}$-wave symmetry. The dispersion for the smaller momenta is remarkably different from that obtained using a constant gap function   \cite{us}.}
\end{minipage}
\end{figure}

The gap function $\Delta(\vk)=0.02\cos(k_x)\cos(k_y)$ is used, which takes on different sign
on the hole and electron FS sheets. The Fermi velocity and momentum in the normal metal are taken as
$p_F=\pi$  and $v_F=1.83$. A potential $\hbar v_FZ\delta(x)$ simulates interface disorder
 \cite{blonder}. For each transverse
momentum (or angle of incidence), the superconducting state conductance, $g_s$, 
may be compared to the normal state conductance, $g_n$.
Figure \ref{gsgn_swave_Z} shows angle resolved relative conductances.  Andreev bound states make
$g_s$ become disorder independent and therefore produce a peak in $g_s/g_n$ for strong disorder, because $g_n$ decreases.
The energy of the ABS depends on the transverse momentum, unlike the usual case of single band 
d-wave superconductors,  where it is zero.   
It has a non monotonic dependence on transverse momentum displayed in Figure \ref{ABSfigure}.

\subsection{Extended s gap symmetry}
 We use the gap function $\Delta(\vk)=0.01\left[\cos(k_x)+\cos(k_y)\right]$ 
which produces nodes on the  electron FS sheets. In contrast with the above, 
we find no  ABS at finite energy for this symmetry. In
principle, the usual zero energy ABS should exist for a different orientation of the
interface relative to the gap function \cite{tanaka}. In Figure 4 we show the relative
conductance.
\begin{figure}[h]
\begin{minipage}{18pc}
\includegraphics[width=18pc]{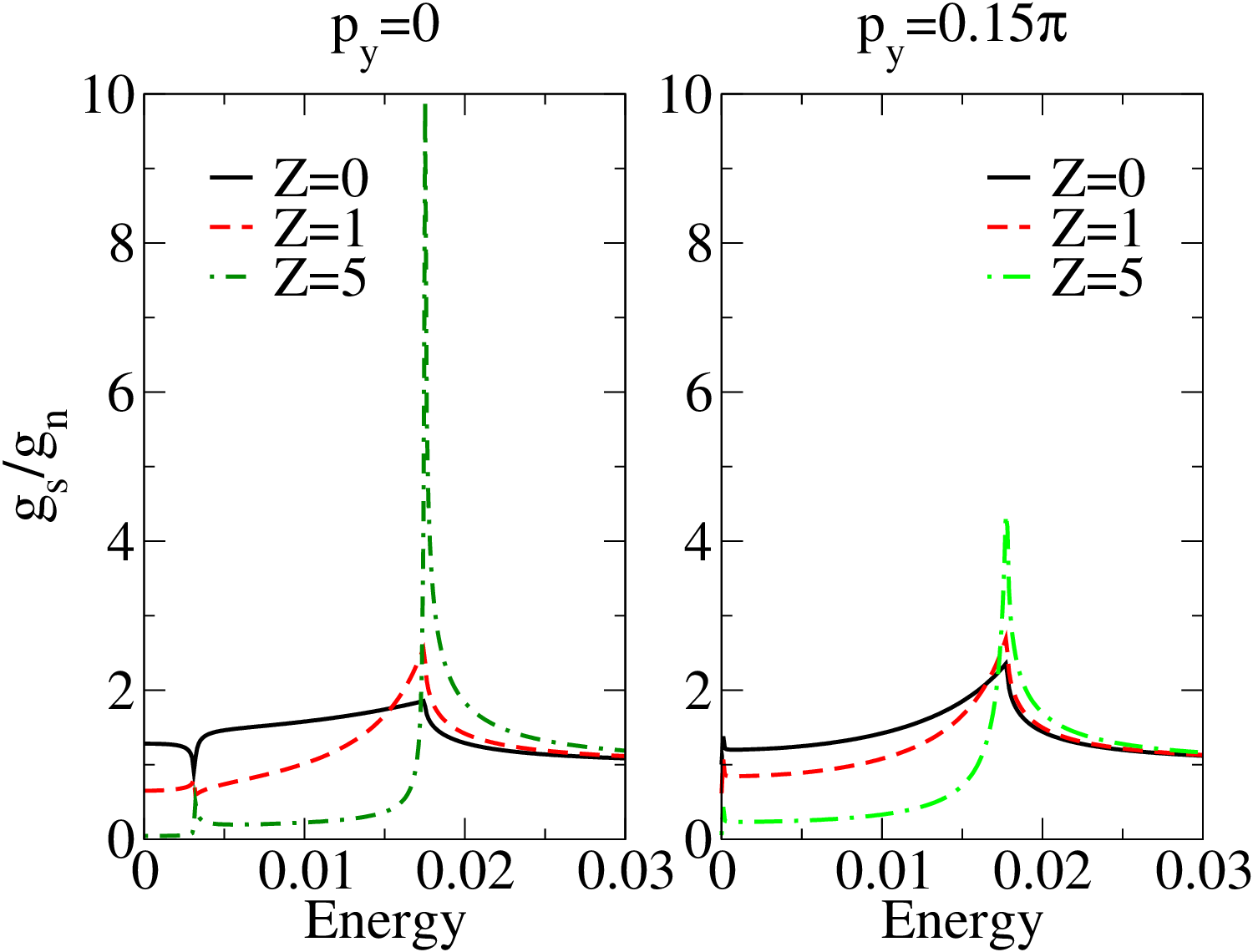}\hspace{2pc}
\caption{\label{grel_dwave}Relative conductance, $g_s/g_n$, for different 
disorder values, $Z$, at normal incidence ($p_y=0)$ and for a finite transverse momentum,
$p_y$,
for extended s-wave symmetry.}
\end{minipage}\hspace{2pc}%
\begin{minipage}{19pc}
\includegraphics[width=19pc]{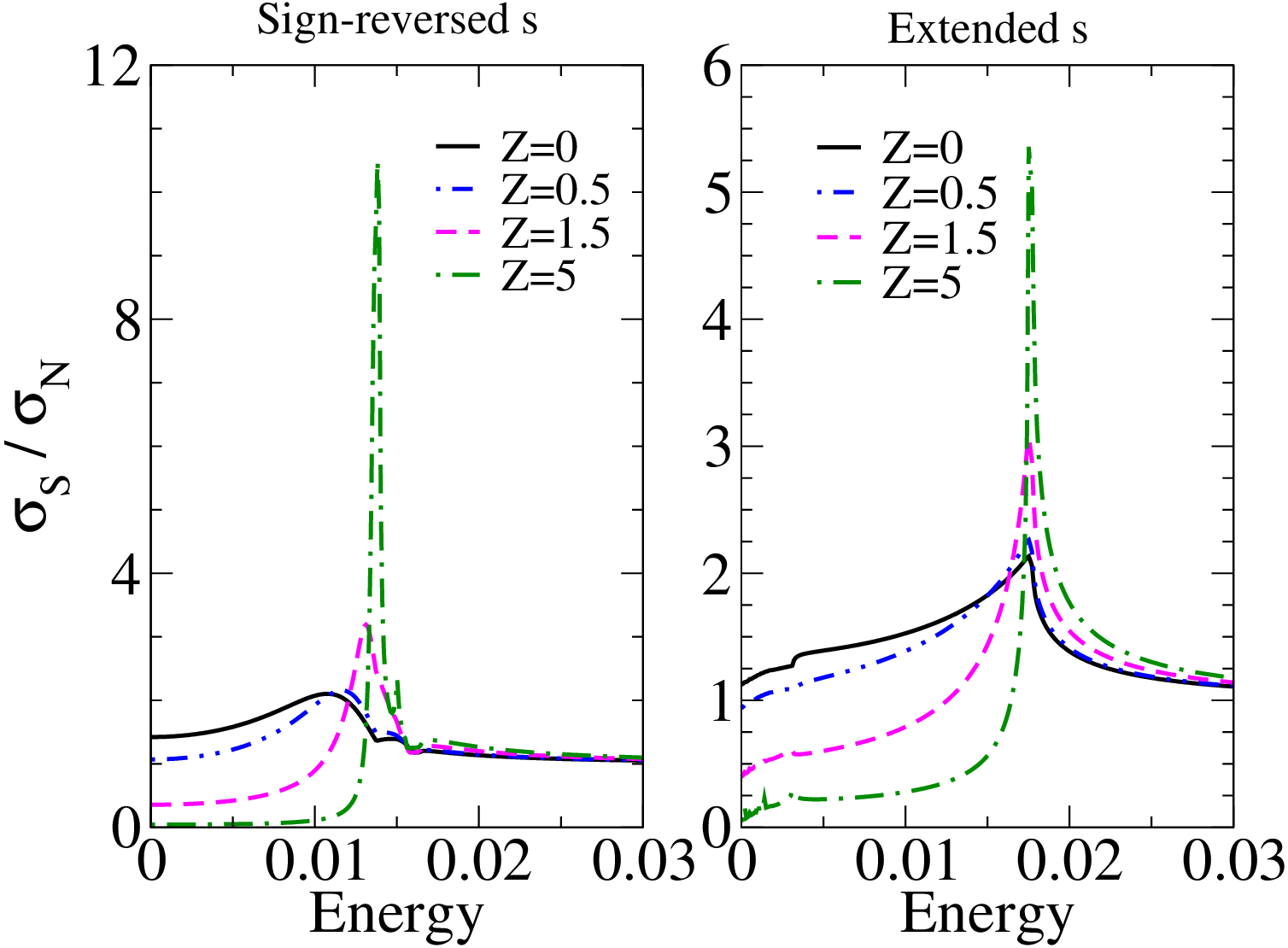}
\caption{\label{grelfull}The integrated relative differential conductance for each gap symmetry.}
\end{minipage} 
\end{figure}

In an experiment, the 
integrated differential conductance over the transverse momenta of incoming particles, 
$\sigma_{S,N}=\int dp_y g_{s,n}$ is observed.
Figure \ref{grelfull} compares  the integrated relative differential conductance 
for different barrier strengths and for the two gap functions above.  



\end{document}